\documentclass[fleqn,usenatbib]{mnras}

\usepackage{newtxtext,newtxmath}
\usepackage[T1]{fontenc}

\DeclareRobustCommand{\VAN}[3]{#2}
\let\VANthebibliography\thebibliography
\def\thebibliography{\DeclareRobustCommand{\VAN}[3]{##3}\VANthebibliography}

\usepackage{graphicx}	% Including figure files
\usepackage{amsmath}	% Advanced maths commands
\title[The 2019 Crab pulsar glitch]{The slow rise and recovery of the 2019 Crab pulsar glitch}

\author[B. Shaw et al.]{B. Shaw,$^{1}$\thanks{E-mail: benjamin.shaw@manchester.ac.uk}
M. J. Keith,$^{1}$
A. G. Lyne,$^{1}$
M. B. Mickaliger,$^{1}$
B. W. Stappers,$^{1}$
J. D. Turner$^{1}$
\newauthor and P. Weltevrede$^{1}$
\\
$^{1}$Jodrell Bank Centre for Astrophysics, School of Physics and Astronomy, University of Manchester, Manchester, UK, M13 9PL\\
}

\date{Accepted XXX. Received YYY; in original form ZZZ}

\pubyear{2021}

\begin{document}
\label{firstpage}
\pagerange{\pageref{firstpage}--\pageref{lastpage}}
\maketitle

\begin{abstract}
We present updated measurements of the Crab pulsar glitch of 2019 July 23 using a dataset of pulse arrival times spanning $\sim$5 months. On MJD 58687, the pulsar underwent its seventh largest glitch observed to date, characterised by an instantaneous spin-up of $\sim$1 $\mu$Hz. Following the glitch the pulsar's rotation frequency relaxed exponentially towards pre-glitch values over a timescale of approximately one week, resulting in a permanent frequency increment of $\sim$0.5 $\mu$Hz.  Due to our semi-continuous monitoring of the Crab pulsar, we were able to partially resolve a fraction of the total spin-up. This \emph{delayed spin-up} occurred exponentially over a timescale of $\sim$18 hours. This is the sixth Crab pulsar glitch for which part of the initial rise was resolved in time and this phenomenon has not been observed in any other glitching pulsars, offering a unique opportunity to study the microphysical processes governing interactions between the neutron star interior and the crust.  
\end{abstract}


%%%%%%%%%%%%%%%%%%%%%%%%%%%%%%%%%%%%%%%%%%%%%%%%%%

%%%%%%%%%%%%%%%%% BODY OF PAPER %%%%%%%%%%%%%%%%%%

\section{Introduction}

The Crab pulsar (B0531$+$21/J0534$+$2200) is the neutron star in the centre of the Crab Nebula, formed in the core-collapse supernova of A.D 1054. It has a rotation frequency of 29.6 Hz (corresponding to one rotation every $\sim$33 ms) and a spin-down rate of \mbox{$-3.7 \times 10^{-10}$ Hz s\textsuperscript{-1}}. The Crab pulsar is one of the youngest neutron stars known and is observable across the electromagnetic spectrum. It is one of the most widely studied pulsars, as the rotational and emission variations it exhibits provide a rich opportunity to understand the structure and radio emission properties of young pulsars.  

The rotation of pulsars is observed to gradually slow down over time due to braking imposed by their strong magnetic fields. In young pulsars, the smooth spin-down may occasionally be interrupted by discrete spin-up events known as glitches. These events are characterised by a sudden increase in the spin frequency of the star which may then decay back towards pre-glitch values, as extrapolated from the pre-glitch model (e.g.,  \citealt{elsk11}, \citealt{ymhj+13}). Additionally, a corresponding jump in the spin-down rate ($\dot{\nu}$) may also be observed. A database of all known pulsar glitches is maintained by the Jodrell Bank Observatory (JBO) and to date lists 577 glitches in 191 pulsars\footnote{http://www.jb.man.ac.uk/pulsar/glitches/gTable.html $-$ accessed 2021 Mar}. In addition, many young pulsars also exhibit a stochastic wandering of their rotational parameters known as \emph{timing noise} (e.g., \citealt{hlk06}). 

Glitches are widely understood to arise from an exchange of angular momentum between an interior ocean of superfluid neutrons and the crystalline crust. The angular momentum of the interior is contained in an array of superfluid vortices. In order to slow down, the superfluid must expel angular momentum though the outward motions of vortices. However, in the inner crust vortices can become pinned to lattice sites, preventing the fluid from slowing \citep{ai75}.  As the crust is braking due to electromagnetic energy losses, a velocity lag grows between the fluid and the crust. Occasionally, a large scale vortex unpinning event occurs, permitting the transfer of some or all of the angular momentum supply into the crust resulting in an observed glitch (see \citealt{hm15} for a r\'esum\'e of glitch models).

The Crab pulsar is one of the most prolific pulsars in terms of its glitch activity, having undergone 30 glitches since its discovery in 1968 (\citealt{ljg+15}; \citealt{sls+18}; \citealt{slb+18}; \citealt{smm+19}). Due to the low and irregular observing cadence at the time, the sample of glitches prior to 1984 may be incomplete.  The range of magnitudes of the spin-frequency increase spans four decades and the inter-glitch intervals are highly irregular (e.g., \citealt{mpw08}). In many of these events $\dot{\nu}$ is also observed to increase and may not completely relax back to pre-glitch values before the next glitch occurrs, resulting in a cumulative effect of the glitches on the long-term spin-down of the star (e.g., \citealt{ljg+15}). 

In most cases, the initial rise in spin-frequency, and any other transient dynamics close in time to a glitch, are too rapid to be resolved. This is, in part due to the finite resources available to observatories coupled with the large numbers of pulsars which are monitored by pulsar timing campaigns, resulting in a low probability that a pulsar will glitch during an observation. However, due to our semi-continuous monitoring of the Crab pulsar at JBO, the rise times of the large glitches of 1989 \citep{lsp92}, 1996 \citep{wbl01} 2004, 2011 \citep{gzl+20} and 2017 \citep{sls+18} were partially resolved in time. The most recent Crab pulsar glitch fortuitously occurred during an observation of the pulsar on 2019 July 23 and was initially reported in  \cite{smm+19}. At the time, the pulsar was still in the post-glitch recovery phase and so the long-term effect of the glitch on the overall spin-down of the pulsar, as well as measurements of any transient components of the glitch were not yet possible.  In this letter, we present updated measurements of the Crab pulsar glitch of July 2019, using an extended dataset comprising $\sim$4 months of post-glitch observations.

\section{Observations}
The Crab pulsar is observed twice daily with the 42-ft telescope at JBO using a bandwidth of 4 MHz centred on 610 MHz. The total daily integration time is typically $\sim$13 hours split into 9 and 4 hour observations. Data are recorded as 1-minute long sub-integrations, folded at the pulse period onto 1024 phase bins, resulting in a time resolution of 0.0330 ms.  We excise radio frequency interference using a median-filter and by manual inspection of individual frequency channels and subintegrations. All data for each observation are finally integrated over the full bandwidth and integration time to form a single pulse profile. An observatory site arrival time (SAT) is computed by cross-correlating the observed profile with a `template' profile that represents an idealised (noise-free) version of the incoming pulse. SATs are converted to barycentric times-of-arrival (TOAs) using the JPL DE436 ephemeris. 

\section{The glitch of 2019 July 23}

\subsection{Timing analysis}

Pulsar glitches are typically identified by the inspection of timing residuals. These are the differences between the observed TOAs from a pulsar and those predicted by an analytical timing model of the pulsar's rotational, astrometric and (where applicable) binary parameters. Where a pulsar's timing model correctly predicts its TOAs, the timing residuals will form a normal distribution about zero.  A glitch is defined as a sudden, discontinuous increase to a pulsar's spin frequency resulting in pulses arriving progressively earlier than the timing model predicts, and this manifests as an increasingly negative departure from zero in a pulsar's timing residuals.

\begin{figure}
    \centering
    \includegraphics[width=1.0\columnwidth]{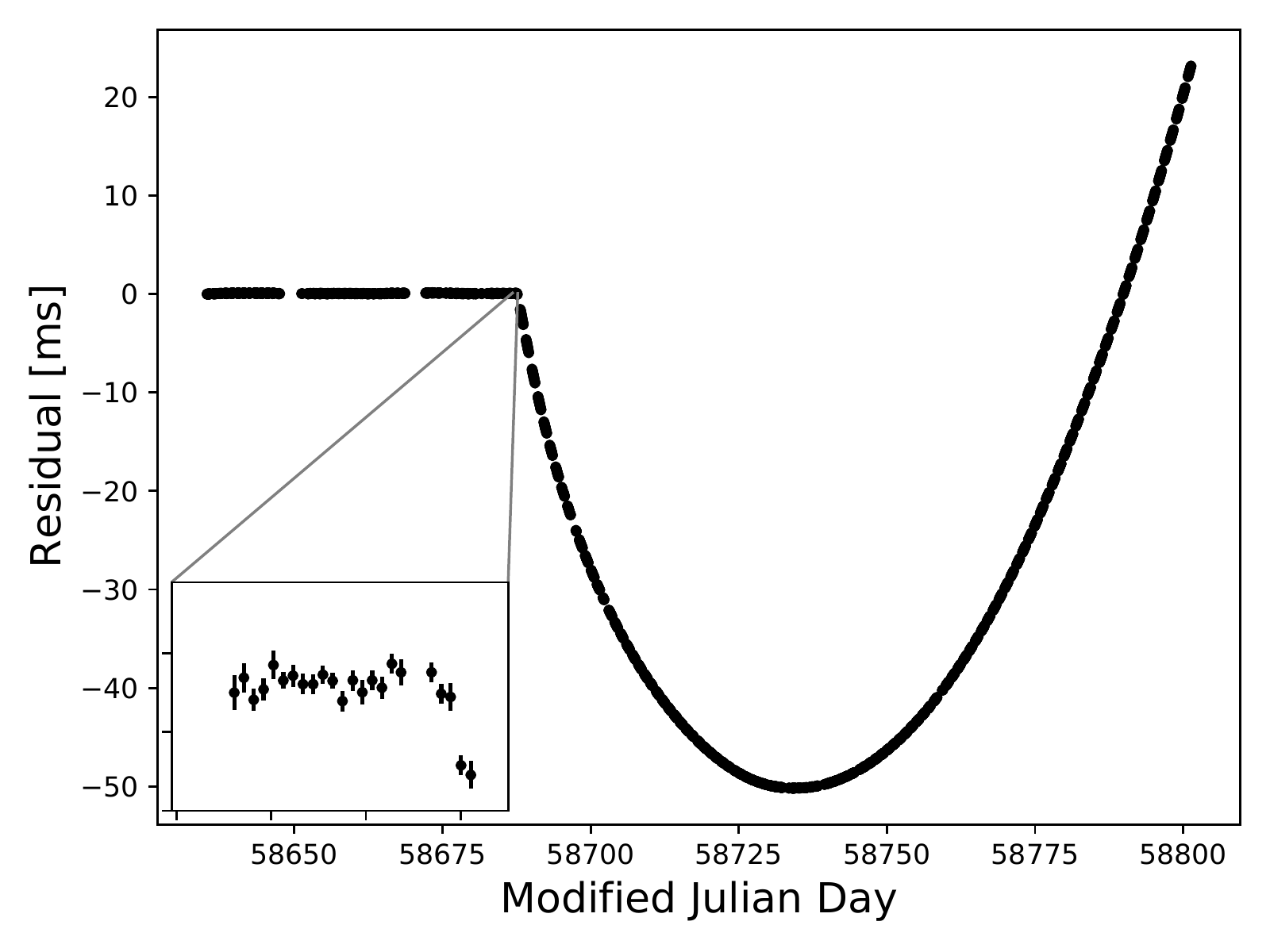}
    \caption{Timing residiuals relative to a simple spin-down model fit to the 52 days of pre-glitch Crab pulsar TOAs. A fit for the pulsar's spin-frequency and spin-down rate, using the position shown in Table \ref{glitch_table}, yields an approximately flat distribution of pre-glitch residuals with values close to zero. The glitch epoch is characterised by the sudden negative departure from zero. From this time onwards, the TOAs are not well described by the pre-glitch model. Each apparent `batch' of timing residuals represents a set of 30 minute observations taken over a single day. The inset shows the individual 30-minute residuals on the day of the glitch.}
    \label{fig:residuals}
\end{figure}

On 2019 July 23 (MJD 58687), the Crab pulsar underwent a moderately-sized glitch during the second and shorter of our two daily observations. In order to measure both the short-term transient components of the glitch as well as any longer term recoveries, we use a dataset comprising  332 TOAs spanning $58635 \leq \mathrm{MJD} \leq 58801$, corresponding to 52 days pre-glitch and 114 days post-glitch. To enhance sensitivity to the shorter term effects of the glitch,  we split the twice-daily TOAs into a larger number of TOAs each of 30 minute duration, thereby improving the observing cadence at the expense of individual TOA uncertainties.  This results in a dataset of 3434 TOAs.  The timing residuals relative to a simple pre-glitch spin-down model are shown in Figure \ref{fig:residuals}. Prior to the glitch, the TOAs were as predicted by the timing model. At the glitch epoch, the pulses begin arriving earlier than predicted by the model. The parabolic nature of the post-glitch residuals suggests that, in addition to the spin-frequency, the spin-down rate of the star has also changed.

\begin{center}
\begin{table}

    \caption{Table of pulsar/glitch parameters.}
    \begin{tabular}{ | p{3.5cm} | p{3.5cm} | }
    \hline
    Parameter & Value \\ \hline
    Right ascension & 05h:34m:31.97s \\
    Declination & +22d:00m:52.07s \\
    Epoch $t_0$ & MJD 58687.5 \\
    Initial $\nu_0$ & 29.616876864(10) Hz \\
    Initial $\dot{\nu}_0$ & $-3.683326(20) \times 10^{-10}$ Hz s\textsuperscript{-1} \\
    Initial $\ddot{\nu}_0$ & $1.98(5) \times 10^{-20}$ Hz s\textsuperscript{-2} \\
    DM & $56.7498(1)$ pc cm\textsuperscript{-3} \\
    DM/dt & $-0.0325(1)$ pc cm\textsuperscript{-3} s\textsuperscript{-1} \\
    \hline
    Glitch epoch $t_\mathrm{g}$ & MJD 58687.565(4) \\
    Permanent spin-up $\Delta \nu_0$ & $5.32(22) \times 10^{-7}$ Hz \\
    Delayed spin-up $\Delta \nu_\mathrm{d}^{(1)}$ & $-3.43(18) \times 10^{-7}$ Hz \\
    Delay time constant $\tau_\mathrm{d}^{(1)}$  & 0.76(7) days \\
    Recovery amplitude $\Delta \nu_\mathrm{d}^{(2)}$ & $7.51(21) \times 10^{-7}$ \\
    Recovery time constant $\tau_\mathrm{d}^{(2)}$ & 6.4(4) days\\ 
    $\Delta \dot{\nu}_0$ & $-1.26(12) \times 10^{-13}$ Hz s\textsuperscript{-1}  \\
    \hline
    EFAC & 1.06(1) \\
    $A_{\mathrm{red}}$ & -9.31(18) \\
    $\alpha$ & 3.82(22) \\
    \hline
    \end{tabular}
    \label{glitch_table}
\end{table}
\end{center}

We model the continuous rotation of the pulsar using a truncated Taylor series,

\begin{equation}
    \label{rot}
    \phi(t) = \phi_0 + \nu_0 (t - t_0) + \frac{1}{2!} \dot{\nu}_0 (t - t_0)^2 + \frac{1}{3!} \ddot{\nu}_0 (t - t_0)^3,
\end{equation}

\noindent where $\phi_0$, $\nu_0$, $\dot{\nu}_0$ and $\ddot{\nu}_0$ are the phase, spin-frequency and its derivatives as measured at the reference epoch $t_0$. The glitch parameters are modelled using an additive function of the form

\begin{equation}
    \label{glitch_pars}
        \begin{aligned}
            \phi_g (t) = &  -\Delta \nu_0 (t - t_g) - \frac{1}{2!}\Delta \dot{\nu}_0 (t - t_g)^2  \\
            & - \sum_i^N \Delta \nu_\mathrm{d}^{(i)} \tau_\mathrm{d}^{(i)}  \left[ 1 - \exp \left( {-\frac{(t - t_g)}{\tau_\mathrm{d}^{(i)}}} \right) \right],
        \end{aligned}
\end{equation}

\noindent whose value is zero if $t \leq t_g$. The first and second terms represent the permanent jumps in the spin frequency and its first derivative after the glitch epoch $t_\mathrm{g}$.  $\Delta \nu_\mathrm{d}^{(i)}$ describes $i$ of $N$ exponentially changing components with a time constant $\tau_\mathrm{d}^{(i)}$. The polarity of $\Delta \nu_\mathrm{d}$ reveals the nature of the transient component of the glitch.  $\Delta \nu_\mathrm{d} > 0$ corresponds to the proportion of the glitch amplitude that exponentially decays after the initial unresolved step. $\Delta \nu_\mathrm{d} < 0$ corresponds to an exponential \emph{rise} after the initial unresolved step. The latter case represents a \emph{slow rise} or \emph{delayed spin-up}. \\

Fitting for glitch parameters was carried out using \textsc{tempo2} \citep{hem06} and the \textsc{enterprise} pulsar timing framework \citep{evt+19}, and  \textsc{ptmcmcsampler} was used to sample over the parameters using a parallel tempering Markov-chain monte-carlo method \citep{ptmcmc}. In addition to the parameters in Equations \ref{rot} and \ref{glitch_pars}, we also include a power-law Gaussian process model for the timing noise observed in the Crab pulsar and an ``EFAC" parameter to model excess white noise on the TOAs due to miscalibrated radiometer noise (see \citealp{lah+13} for a description of the noise model used). We fit for the index, $\alpha$, and amplitude, $A_\mathrm{red}$, of the power-law Gaussian process such that the power-spectral density of the red noise at a fluctuation frequency $f$ is given by

\begin{equation}
    P(f) = \frac{A_\mathrm{red}^2}{12\pi^2} \left(\frac{f}{\mathrm{yr}^{-1}}\right)^{-\alpha} \mathrm{yr}^3.
\end{equation}

\noindent All parameters were determined using an uninformative uniform prior, and we analytically marginalise over $\phi_0$, $\nu_0$, $\dot{\nu}_0$ and $\ddot{\nu}_0$.

The exact functional form of Equation \ref{glitch_pars} that best models a glitch cannot be known \emph{a-priori}. For example, multiple exponential terms may be required to model the post-glitch transient behaviour of the pulsar. The Vela pulsar for instance, has been shown to recover from glitches over multiple distinct timescales (e.g., \citealt{dlm07}). For this reason we attempted to model the glitch with up to three exponential terms, however we reject the zero- and one-exponential models as the maximum-likelihood solutions still show significant transient events at the glitch epoch. The three-exponential model did not offer any improvement over two-exponential model.  Therefore, we only consider the two-exponential model for the remainder of the paper.

\begin{figure*}
    \centering
    \includegraphics[width=2.0\columnwidth]{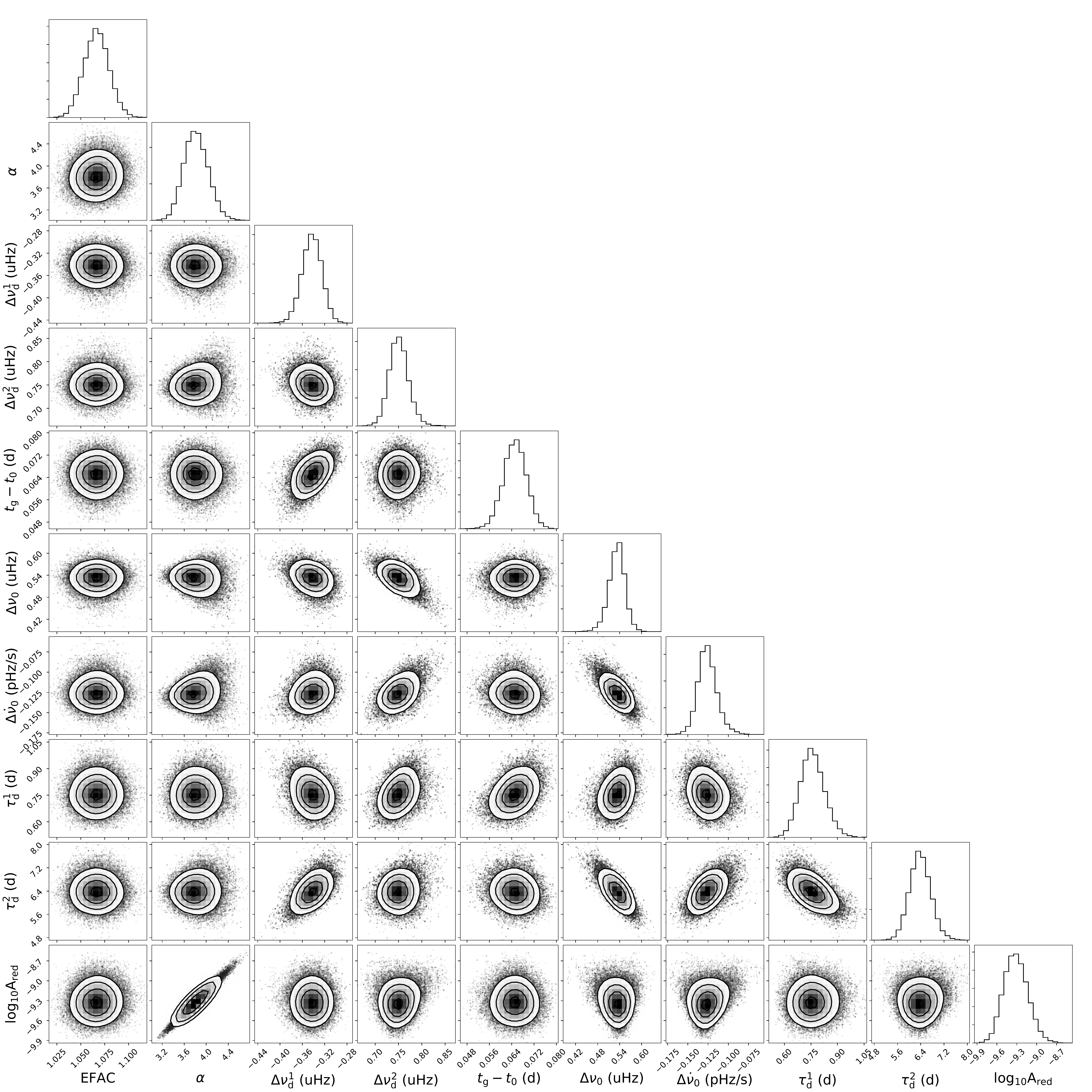}
    \caption{Univariate and bivariate posterior probability distributions for the glitch parameters and the red and white noise parameters. The contours represent the 1-, 2- and 3-$\sigma$ confidence intervals for each pair of parameters. The glitch epoch posterior distribution is shown relative to the the period epoch $t_0$ in the timing model (see Table \ref{glitch_table}).}
    \label{fig:corner}
\end{figure*}

By fitting for the glitch parameters using the two-exponential model, we estimate a permanent step change in the spin-frequency $\Delta \nu_0 = 5.32 \pm 0.22 \times 10^7$ Hz, corresponding to a fractional change in the spin-frequency \mbox{$\Delta \nu_0 / \nu_0=(18.0\pm0.8)\times10^{-9}$}. We measure the amplitude of the first transient component to be $\Delta \nu_{\mathrm{d}}^{(1)}=(-3.43\pm0.18)\times10^{-7}$ Hz, rising exponentially over \mbox{$\tau_{\mathrm{d}}^{(1)} = 0.76 \pm 0.07$ days} and the second transient component to be  \mbox{$\Delta \nu_{\mathrm{d}}^{(2)}=(7.51\pm0.21)\times10^{-7}$ Hz},  decaying exponentially over \mbox{$\tau_{\mathrm{d}}^{(2)}=6.4\pm0.4$ days} . The former term corresponds to a \emph{slow rise} in frequency at the glitch epoch that was partially resolved in time. The combinations of these amplitudes corresponds to a total spin-up of $\Delta \nu_{\mathrm{total}}=(9.4\pm0.4)\times10^{-07}$ Hz (a fractional total spin-up of $\Delta \nu_{\mathrm{total}} / \nu_0=(31.7\pm1.2)\times10^{-9}$) making this event the seventh largest glitch observed in the Crab pulsar.  We also measure a permanent spin-down change $\Delta \dot\nu_0=(-1.26\pm0.12)\times10^{-13}$ Hz s\textsuperscript{-1}. These results are summarised in Table \ref{glitch_table}. We show the one- and two-dimensional posterior distributions of the glitch, red, and white noise parameters in Figure \ref{fig:corner}. After adding the measured glitch parameters to the timing model, the resulting timing residuals have a RMS value of $\sim10\,\mu{\rm s}$, corresponding to $\sim$0.03\% of one pulse period.

\begin{figure}
    \centering
    \includegraphics[width=1.0\columnwidth]{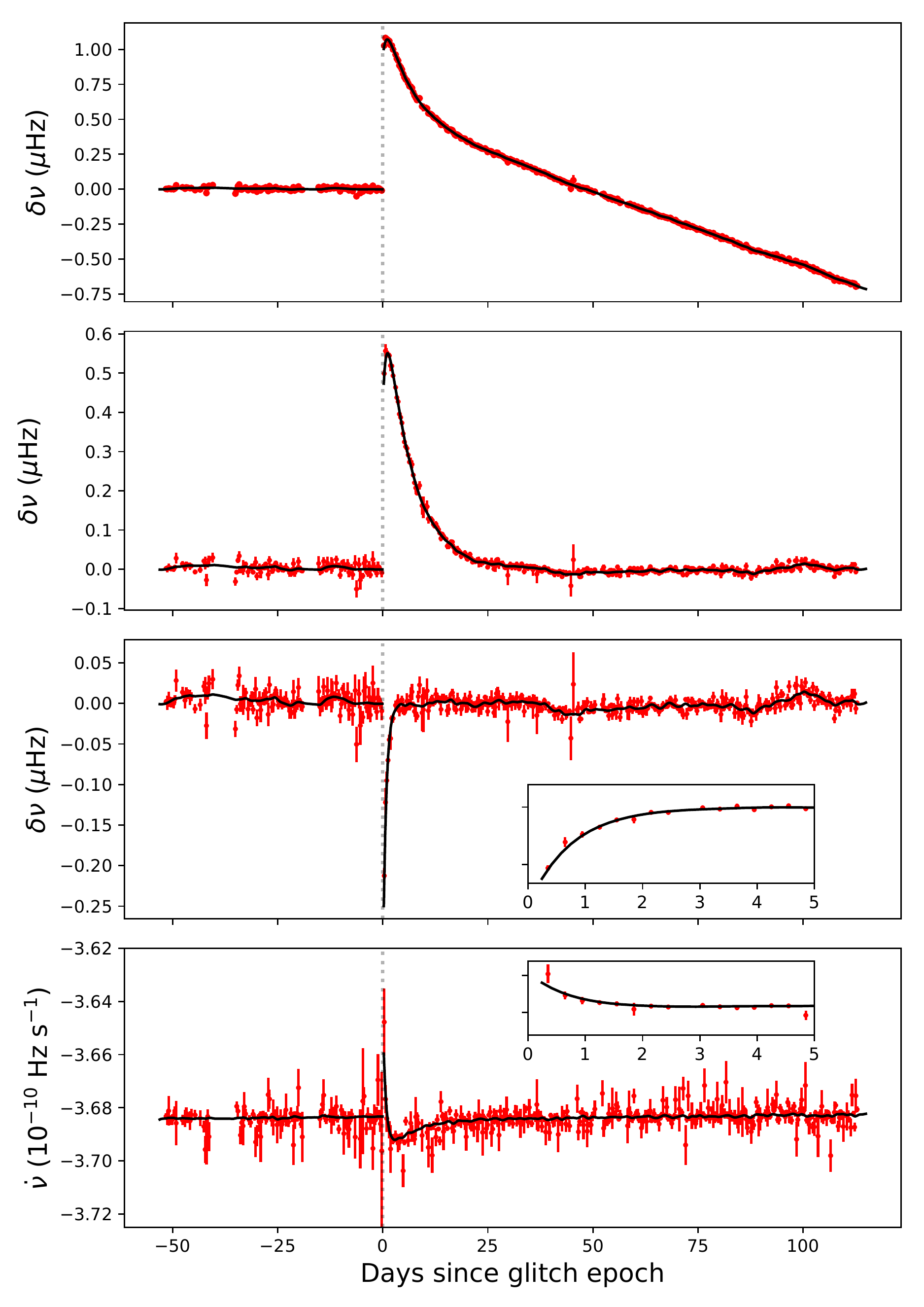}
    \caption{The evolution of the rotation of the Crab pulsar close to the epoch of the July 2019 glitch (vertical dotted line). Top panel: The frequency residuals $\delta \nu$ relative to the extrapolation of the spin-frequency prior to the glitch and after subtracting the pre-glitch spin-down rate. Second panel: Frequency residuals after subtraction of the permanent step changes in the spin rate $\Delta \nu_0$ and spin-down rate $\Delta \dot{\nu}_0$. The slow rise is clearly visible as the steep rise in $\delta \nu$ immediately after the glitch epoch.  Third panel: As above after subtraction of the positive exponential (recovery) term $\Delta \nu_\mathrm{d}^{(2)}$ leaving only the negative exponential (slow rise) term $\Delta \nu_\mathrm{d}^{(1)}$. Lower panel: Evolution of the spin-down rate. A downward deflection represents in increase in the magnitude of $\dot{\nu}$. The red data points were computed directly from the TOAs using a striding boxcar method. The black lines represent the rotational evolution generated from an analytical model of the glitch (See text). The insets show the rotational evolution in the first five day after the glitch.}
    \label{fig:nu_nudot}
\end{figure}

The  spin frequency residuals and the evolution of the spin-down rate around the 2019 glitch are shown in Figure \ref{fig:nu_nudot}. The black lines represent an analytical model of the frequency evolution generated from the maximum likelihood values of the glitch and red noise parameters. The top panel shows the residuals relative to the pre-glitch frequency. The delayed spin-up is discernible as the short period of frequency increase immediately following the measured glitch epoch (dashed line).  The second panel shows the frequency residuals following subtraction of the permanent step changes $\Delta \nu_0$ and $\Delta \dot{\nu}_0$, leaving behind only the two exponential terms. The third panel shows the residuals following the subtraction of the positive exponential recovery term, leaving only the negative exponentially rising component. The evolution of the spin-down rate (lower panel) shows the recovery of the spin-down towards pre-glitch values occurring until $\sim$15 days after the glitch, preceded by a strong negative slope immediately after the glitch, corresponding to the slow-rise in spin-frequency.  We use the striding boxcar method described in \cite{sls+18} to examine the spin-frequency and spin-down rate evolution from the 3434 TOAs directly. We use a boxcar window of width 1.5 days and a step size of 0.3 days ($\sim$7 hours), fitting Equation \ref{rot} to the TOAs within each window. The results are shown as the red points in Figure \ref{fig:nu_nudot} and are in good agreement with analytical model of the glitch parameters.

\section{Discussion}

We have presented an updated measurement of the Crab pulsar glitch of 2019 July 23. An initial estimate of the spin-frequency increment was determined just two days after the onset of the glitch \citep{smm+19}, such that the long-term effects of the glitch and any post-glitch relaxation could not be accurately measured. In this work we present measurements of the long-term change to the rotational parameters of the pulsar after the glitch, and we also detect two exponential transient components - one corresponding to post-glitch relaxation occurring on a timescale of $\sim$6 days and a second, slowly-rising \emph{delayed spin-up} component occurring over $\sim$0.7 days.  The delayed spin-up is the sixth known example of such a phenomenon, having also occurred in the large Crab pulsar glitches of 1989 (0.8 days), 1996 (0.5 days), 2004 (1.7 days), 2011 (1.6 days) and 2017 (1.7 days). These rise times exceed those expected by fluid-crust coupling models (e.g., \citealt{als84}; \citealt{gca18}).  Delayed spin-ups have so far only been observed as components of larger Crab pulsar glitches. It may be the case that such events are too small and/or brief to be resolved near smaller glitches unless such glitches are 'caught in the act' during an observation.  We note however, that no delayed spin-up was identified near the small Crab pulsar glitch of 1997 \citep{wbl01}, despite it occurring just 1 hour prior to the commencement of an observation.  

The occurrence of the delayed spin-ups in the Crab pulsar has been the focus of a number of studies, as they offer a unique insight into the microphysical interactions of glitches as well as the physical location within the star where they are triggered.  \cite{accp96} and \cite{geaa19} showed that such delays could arise due to vortices moving inwards towards the neutron star core during unpinning events, mediated by motions in the stellar crust. They have also been attributed to the deposition of energy into the crust by starquake events \citep{ll02},  and to the formation of vortex sheets \citep{kh18}.  \cite{hkaa18} developed a hydrodynamical model of pulsar glitches and determined that Crab pulsar glitches likely originate in the crust with relatively low levels of mutual friction between the crust and core giving rise to delayed spin-ups, in contrast with Vela glitches which are likely triggered in the core. More recently \cite{sc20} showed that the pinning of vortices to \emph{fluxoids} could affect the mutual friction, giving rise to delayed spin-ups. 

To-date, delayed spin-ups have not been detected in any other glitching pulsars, although measurements of very short-term transient behaviours such as these are made possible due to our high-cadence and long dwell time observations of the Crab pulsar. Therefore, delayed spin-ups may be more widespread throughout the glitching pulsar population. In contrast, the Vela pulsar is monitored near-continuously by Southern hemisphere observers, and rise times have been constrained to within 5 s \citep{pdh+18}, 30 s \citep{dlm07} and 40 s \citep{dml02} and no slow rises have been observed. Determining which, if either, of these cases is typical of the glitching pulsar population relies on a larger sample of young pulsars being afforded similar telescope resources to the Vela and Crab pulsars. Next-generation telescopes have a key role to play in this respect. For example, with the wide field-of-view and large number of tied-array beams of the Square Kilometre Array (SKA1-LOW), it will be possible to time more pulsars, more frequently (e.g., \citealt{wre+15}), allowing recoveries and other transient rotational irregularities associated with glitches to be studied in more detail.  

\section{Conclusions}

We have presented an updated measurement of the Crab pulsar glitch of 2019 July 23 using $\sim$5 months of data. We measure a total instantaneous spin-up of $\sim1 \mu$Hz corresponding to a fractional instantaneous frequency change of $\sim32 \times 10^{-9}$. Following a period of recovery, the pulsar's spin frequency was increased by $\sim$50\% of the total spin-up. We also measured a significant change to the spin-down rate following the glitch. Due to our semi-continuous monitoring the Crab pulsar, part of the initial spin-up was resolved in time, occurring over a timescale of $\sim$18 hours. This $\emph{delayed spin-up}$ is the sixth detected such event, having only occurred previously in other medium to large Crab pulsar glitches. We continue to monitor the Crab pulsar for new glitch events.  

%%%%%%%%%%%%%%%%%%%%%%%%%%%%%%%%%%%%%%%%%%%%%%%%%%
\section*{Data Availability}

The data underlying the work in this letter are available upon reasonable request.

\section*{Acknowledgements}
Pulsar research at Jodrell Bank is supported by a consolidated grant from the UK Science and Technology Facilities Council (STFC). 



\bibliographystyle{mnras}
\bibliography{journals,psrrefs,modrefs} % if your bibtex file is called example.bib

%\begin{thebibliography}{99}
%\end{thebibliography}

%%%%%%%%%%%%%%%%%%%%%%%%%%%%%%%%%%%%%%%%%%%%%%%%%%

%%%%%%%%%%%%%%%%% APPENDICES %%%%%%%%%%%%%%%%%%%%%

%\appendix



%%%%%%%%%%%%%%%%%%%%%%%%%%%%%%%%%%%%%%%%%%%%%%%%%%

% Don't change these lines
\bsp	% typesetting comment
\label{lastpage}
\end{document}